\documentclass [preprint]{revtex4-1} 
\usepackage{amssymb}
\usepackage{graphicx}
\usepackage{bm}
\usepackage{textcomp}
\usepackage{amsmath}
\usepackage{natbib}
\usepackage{units}
\usepackage{color}
\usepackage{epstopdf}

\begin{document}

\title{Thermal Gating of Charge Currents with Coulomb Coupled Quantum Dots}

\author{H.~Thierschmann$^1$, F. Arnold$^1$,  M.~Mitterm\"uller$^1$, L.~Maier$^1$, C.~Heyn$^2$, W.~Hansen$^2$, H.~Buhmann$^1$, and L.~W.~Molenkamp$^1$ }

\affiliation{ $^1$Physikalisches Institut (EP3), Universit\"at W\"urzburg, Am Hubland, D-97074, W\"urzburg, Germany\\
$^2$Institute of Applied Physics, University of Hamburg, Jungiusstrasse 11, D-20355 Hamburg, Germany\\
}

\date\today

\begin{abstract}
We have observed thermal gating, i.e. electrostatic gating induced by hot electrons. The effect occurs in a device consisting of two capacitively coupled quantum dots. The double dot system is coupled to a hot electron reservoir on one side (QD1), whilst the conductance of the second dot (QD2) is monitored. 
When a bias across QD2 is applied we observe a current which is strongly dependent on the temperature of the heat reservoir. This current can be either enhanced or suppressed, depending on the relative energetic alignment of the QD levels. Thus, the system can be used to control a charge current by hot electrons.
\end{abstract}


\maketitle

In recent years thermoelectrics, thermionics and thermal management in small-scaled devices have become important subjects both in basic and applied solid-state research \cite{Giazotto2006, Shakouri2011}. In order to control the heat flow on the nanometer scale, a number of promising concepts have been proposed and partly realized, including solid-state thermal rectifiers \cite{Terraneo2002, Chang2006, Scheibner2008, Matthews2012, Tseng2013}, thermal transistors \cite{Yigen2013, Ben2014}, and nano-refrigerators \cite{Prance2009, Whitney2013}. Moreover, new concepts for highly efficient energy harvesting devices have recently been proposed which use a system of two quantum dots (QDs) as a central building block to convert heat into a directed current \cite{Sanchez2011, Sothmann2012, SothmannB2012}. The key feature of these systems is the capacitive inter-dot coupling \cite{Molenkamp1995} which enables energy exchange between the QDs while particle transfer is blocked.
Here we show how a system of two Coulomb coupled QDs acts as a thermal gate for charge currents.
One of the dots (QD1) can exchange electrons with a hot reservoir only. The other dot (QD2) connects two reservoirs of equal but lower temperature. If a small potential difference is applied across QD2, we observe that the resulting current can be either enhanced or suppressed by variation of the temperature in the hot bath connected to QD1. An intuitive picture is given which explains the underlying mechanism. It is shown that this effect of thermal gating is in fact strongly related to correlations between fluctuations in the occupation number of both QDs.

\begin{figure}[]
 	\centering
 		\includegraphics[width=0.8\linewidth]{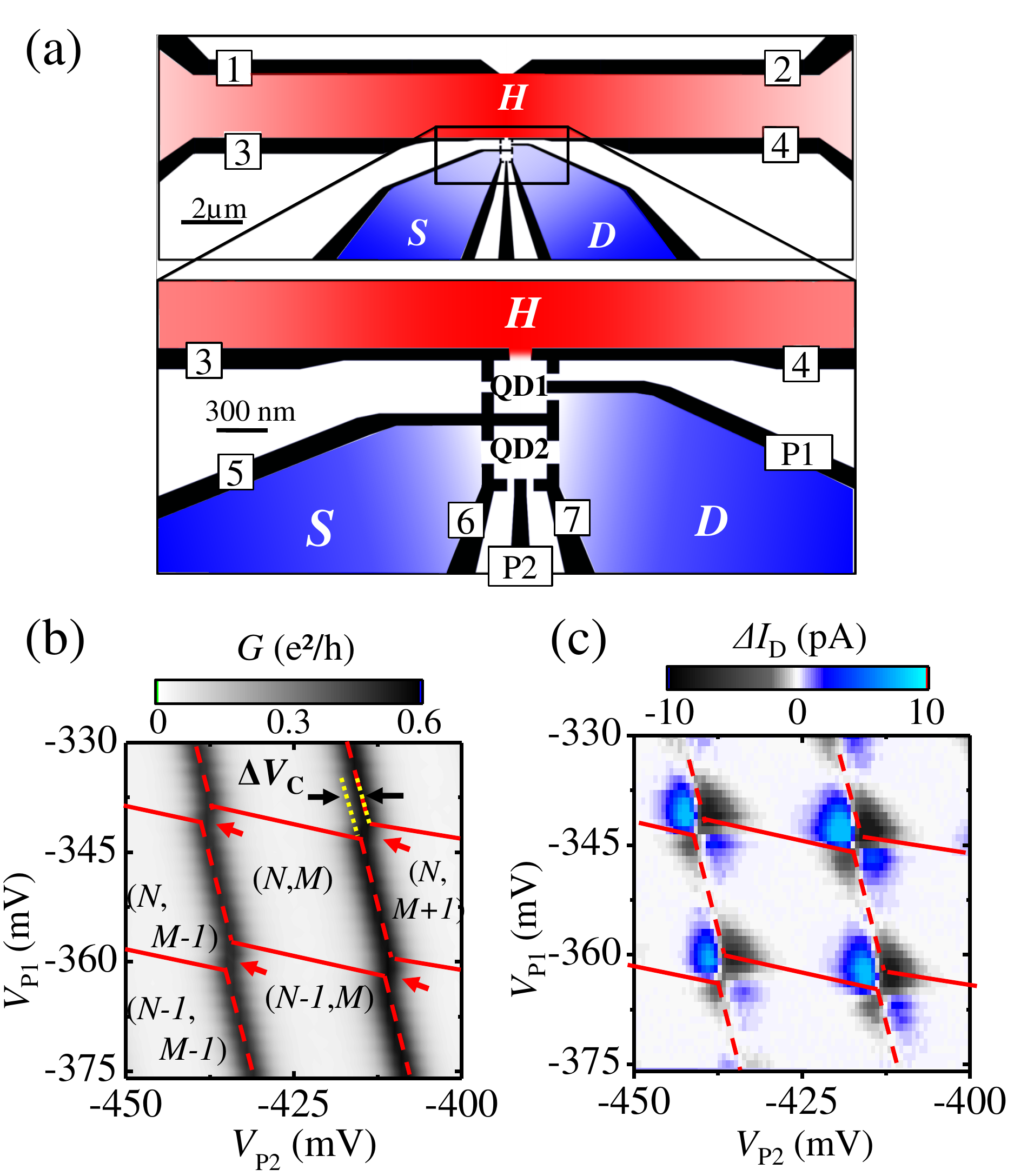}
 		\caption{
 		 \textbf{(a)} Schematic design of the gate structure (black). Gates are labeled with numbers 1-7, P1 and P2. Electronic reservoirs are denoted S, D (both blue) and H (red). \textbf{(b)} Stability diagram of the QD-system showing the conductance of QD2. The characteristic honeycombs are indicated with red lines. QD occupation numbers are denoted with $N$, $M$. $\Delta V_\text{C}$ indicates the capacitive coupling energy. \textbf{(c)} Current signal $\Delta I_{\rm D}$ in reservoir D with $V_{\rm SD} \approx \unit[-30]{\mu V}$ for $T_{\rm H} \approx T_{\rm S,D} + \unit[100]{mK}$. 
 		}
 	\label{Sample}
 \end{figure}

The device is processed by means of optical and e-beam lithography and subsequent metalization of Ti/Au-electrodes (gates) on a GaAs/AlGaAs heterostructure with a two-dimensional electron gas (2DEG) 94 nm below the surface (carrier density $n = 2.4 \times 10^{11}$ cm$^{-2}$, electron mobility $\mu = 0.69 \times 10^6$ cm$^2$/Vs).
The gate structure is shown in Fig.~\ref{Sample}~(a). The gates (black) are labeled 1 through 7, P1 and P2. The electron reservoirs are denoted S, D and H.
The QDs are defined by gates 3 to 7, labeled QD1 and QD2. QD1 is in direct contact with reservoir H (red), which will serve as a hot electron reservoir. QD2 is connected to reservoirs S and D (both blue) which are at a lower temperature, representing the conductor circuit. The junctions are tuned into the tunneling regime by adjusting the voltages of the appropriate gates. The plunger-gate voltages P1 and P2 control the chemical potentials $\mu^{(1)}$ and $\mu^{(2)}$ of the QDs such that the electron occupation numbers can be adjusted individually. The two QDs are separated by an electrostatic barrier (gate 5), which suppresses the electron transfer between the dots. At the same time, QD1 and QD2 interact electrostatically due to the small spatial separation. This means that the occupation number of each QD affects the chemical potential of the other QD, and thus $\mu^{(1)} = \mu^{(1)}(N,M)$ [$\mu^{(2)} = \mu^{(2)}(N,M)$] where $N$ and $M$ are the occupation numbers of QD1 and QD2, respectively \cite{Molenkamp1995, VanderWiel2003}.
Reservoir H defined by gates 1, 2, 3 and 4 forms a channel of 20 $\mu$m length and 2 $\mu$m width. Opposite QD1, a constriction is created by gates 1 and 2 which can be used as a voltage probe in the channel. Its conductance is set to $G = \unit[10]{e^2/h}$ by adjusting the gate voltages $V_1$ and $V_2$ thus ensuring that no thermovoltage is created across this junction when the temperature in reservoir H is increased \cite{Molenkamp1990}.

The sample is mounted in a top-loading dilution refrigerator with base temperature $T_\text{base} =  \unit[80]{mK}$. 
For a conductance characterization of QD2 with all reservoirs at $T_{\rm base}$, an excitation voltage $V_\text{ac}= 5~\mu\text{V}$ ($f=\unit[113]{Hz}$) is applied between reservoirs S and D. The current is measured with a lock-in, using a current amplifier that connects D to a virtual ground potential. By varying $V_\text{P1}$ and $V_\text{P2}$ one obtains the so-called stability diagram of the QD-system \cite{VanderWiel2003}, shown in Fig.~\ref{Sample}~(b) where the conductance $G$ of QD2 is displayed in a gray scale as a function of the voltages $V_\text{P1}$ and $V_\text{P2}$.

Along the horizontal axis $V_\text{P2}$, we observe two conductance resonances which identify those gate voltage configurations for which $\mu^{(2)}$ is aligned with $\mu_{\rm S}$ and $\mu_{\rm D}$. They are separated by the Coulomb charging energy of QD2.
Due to the mutual capacitive coupling the energetic position of $\mu^{(2)}$ is affected by the energy of QD1. This leads to a continuous shift of the conductance resonances for larger $V_\text{P1}$ towards smaller $V_\text{P2}$ [dashed, red lines in Fig.~\ref{Sample}~(b)]. When $\mu^{(1)}$ aligns with $\mu_\text{H}$, $N$ changes by one [solid, red lines in Fig.~\ref{Sample}~(b)]. This causes discrete jumps for the conductance resonances of QD2, indicated by red arrows in Fig.~\ref{Sample}~(b). These jumps are a result of the capacitive inter dot coupling which leads to the transfer of the energy $E_\text{C}$: $\mu^{(2)}(N+1, M) = \mu^{(2)}(N,M) + E_\text{C}$ \cite{Molenkamp1995, VanderWiel2003}.
Hence, the charge occupation numbers of both QD1 and QD2 are stable only in the regions enclosed by solid and dashed lines in Fig.~\ref{Sample}~(b). The energy $E_\text{C}$ can be calculated from the displacement of the conductance resonance along the $V_\text{P2}$-direction, $\Delta V_{\rm C}$, indicated by yellow dotted lines in the figure.
Using the gate efficiency $\alpha_2=0.032$ obtained from d$I$/d$V$ characterization of QD2 yields $E_\text{C}\approx \unit[90]{\mu V}$.

In order to subject the QD-system to a temperature difference, we make use of a current heating technique \cite{Molenkamp1990}: An ac-current $I_\text{h} = \unit[150]{nA}$ with frequency $f = \unit[113]{Hz}$ is applied to the heating channel (reservoir H). Because of the strongly reduced electron-lattice interaction in GaAs/AlGaAs 2DEGs at low temperature, the energy is dissipated into the lattice only in the wide contact reservoirs. On a length scale of a few $\mu$m, however, electron-electron scattering dominates electron-phonon scattering, resulting in a thermalized hot Fermi distribution of the electrons in the channel only.
Based on QPC-thermometry \cite{Molenkamp1990} we estimate that for a current of $I_\text{h} = 150~\text{nA}$, $T_\text{H}$ increases by $\Delta T \approx 100~\text{mK}$. The ac-heating causes the temperature in the heat reservoir to oscillate at $2f = \unit[226]{Hz}$ between $T_\text{base}$ and $T_\text{max} = T_\text{base} + \Delta T $. This ensures that all temperature-driven effects also oscillate at frequency $2f$, enabling straight forward lock-in detection.
Next, a dc-voltage source is connected to S which applies $V_\text{S,GND}$ while the current amplifier (input impedance $R_\text{imp}= \unit[2]{k \Omega}$) connecting D to ground potential is read out by a lock-in amplifier detecting at $2f=226~\text{Hz}$. This allows us to determine the change of the current in the drain contact $\Delta I_\text{D}$ due to variation of $T_\text{H}$. 

\begin{figure}[]
	\centering
		\includegraphics[width=0.8\linewidth]{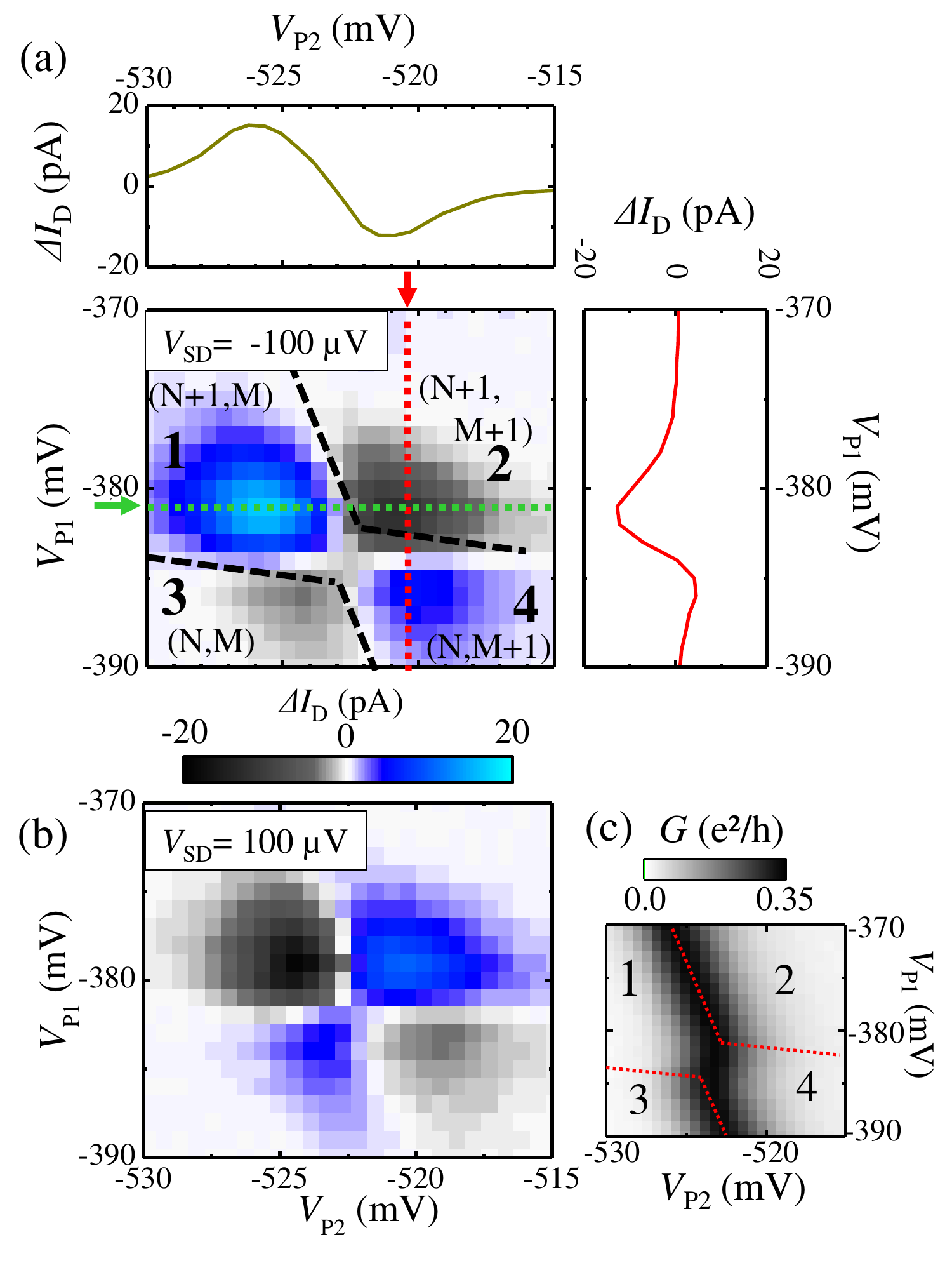}
		\caption{\textbf{(a)} Color scale plot of the change of current $\Delta I_\text{D}$ with $T_\text{H}$ in the region of the honeycomb vertex. Black, dashed lines indicate the boundaries of the stability regions denoted 1-4. Data taken for $V_\text{SD} = -100~\mu$V. Top and side panel show single traces extracted for $V_\text{P1} = -381~\text{mV}$ and $V_\text{P2} = -521~\text{mV}$, indicated by horizontal (green) and vertical (red) line in the color scale plot \textbf{(b)} Data taken for inverted bias voltage: $V_\text{SD} = 100~\mu\text{V}$. \textbf{(c)} Conductance stability diagram for the same gate voltage region.}
	\label{Butterfly}
\end{figure}

With $V_\text{S,GND} \approx \unit[-30]{\mu V}$ we obtain the data shown in Fig.~\ref{Sample}~(c). The lines delimiting the stability regions are indicated.
Surrounding each stability region vertex we observe a four-leaf clover shaped structure that is composed of positive and negative current changes of up to $\pm \unit[8]{pA}$. The sign changes occur at the transitions from one quarter of a ``clover leave'' to the adjacent ones. Diagonally opposite regions exhibit identical sign.
A closeup of a similar clover-leaf structure, obtained for slightly different values for $V_\text{P1}$ and $V_\text{P2}$, is given in Fig.~\ref{Butterfly}~(a) for $V_\text{S,GND}=\unit[-100]{\mu V}$. [For the measurements shown in Fig.~\ref{Butterfly} the current amplifier is replaced by a resistor $R = \unit[100]{k \Omega}$, the voltage drop across which is detected by the lock-in at $2f$.]
The corresponding conductance stability vertex is shown in Fig.~\ref{Butterfly}~(c). A direct comparison identifies the four parts of the clover-leaf pattern with different stability regions of the vertex: Sections 1 ($N$+1, $M$) and 4 ($N$, $M$+1) produce a positive signal while for sections 2 ($N$+1, $M$+1) and 3 ($N$, $M$) negative $\Delta I_\text{D}$ are observed.
A single trace extracted from the color scale plot for constant $V_\text{P1} = \unit[381]{mV}$ (green, horizontal line) is shown in the top panel of Fig.~\ref{Butterfly}~(a). It exhibits a maximum and a minimum at $V_{\rm PG}$ corresponding to dot occupation ($N$+1, $M$) and ($N$+1, $M$+1), respectively. In between, the signal changes approximately linearly with $V_{\rm P2}$. Moving further away from the vertex causes the signal to decay. A trace extracted along the $V_\text{P1}$ axis for constant $V_\text{P2} = \unit[521]{mV}$ (red, vertical line) behaves likewise (side panel).
In a next step the dc-voltage applied to S is reversed, so that $V_\text{S,GND} = \unit[100]{\mu V}$. The result is given in Fig.~\ref{Butterfly}~(b). Clearly, the clover leaf pattern is reproduced, however, with all signs inverted.

\begin{figure}[]
	\centering
		\includegraphics[width=0.7\linewidth]{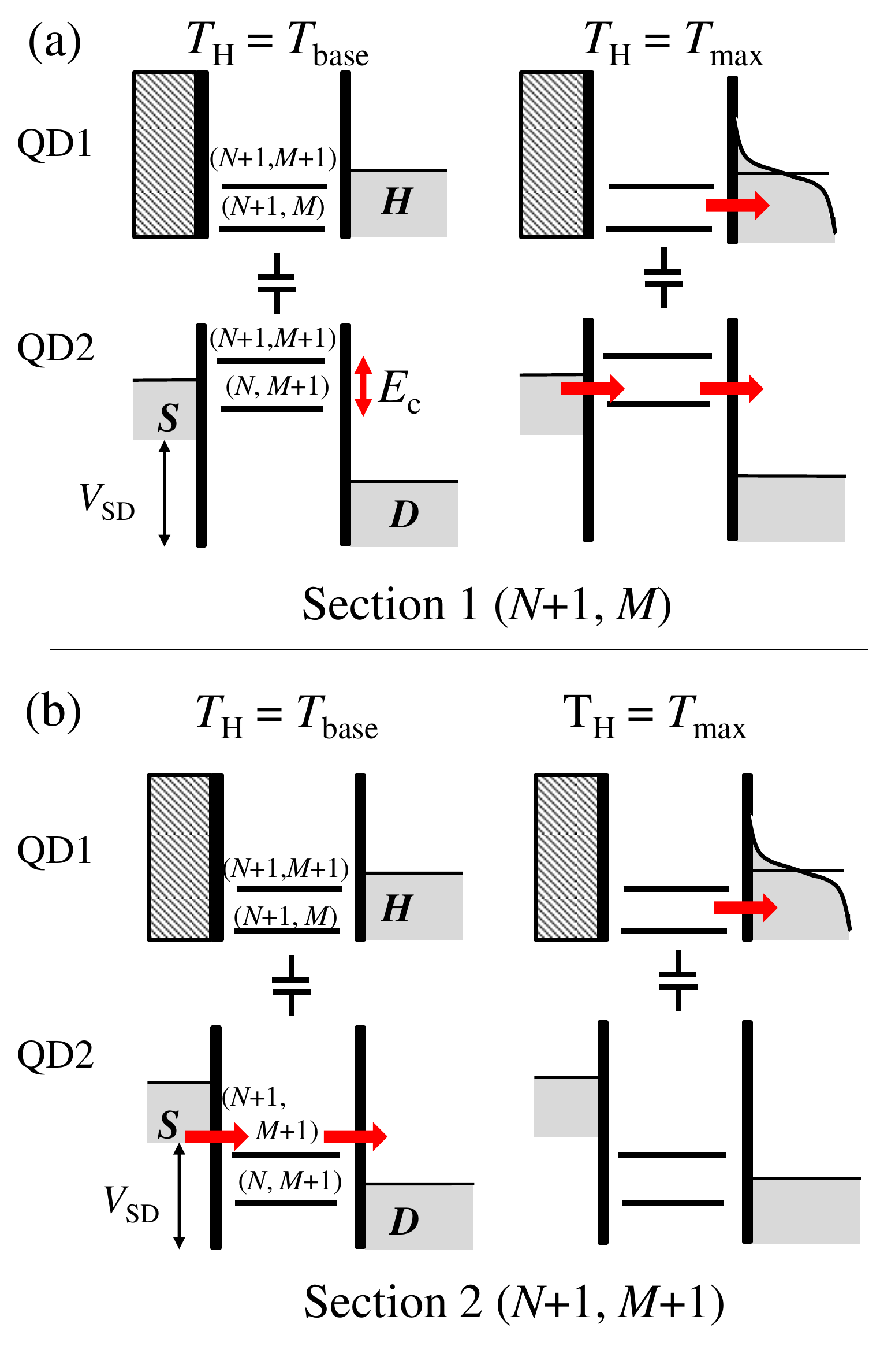}
		\caption{
		Schematic energy diagram of the QD-system showing the alignment of chemical potentials for stability regions \textbf{(a)} ($N+1,M$), section 1 and \textbf{(b)} ($N+1,M+1$), section 2. Each configuration is shown for low temperature (left) and high temperature (right) in reservoir H and with finite $V_{\rm SD}$. Red arrows indicate enhanced occupation fluctuations.   
		}
	\label{Cartoon}
\end{figure}

We now discuss qualitatively how we can understand this behavior.
As is evident from Fig.~\ref{Butterfly}, the sign of $\Delta I_\text{D}$ does not change over a single stability section of the system [labeled 1 - 4 in Fig.~\ref{Butterfly}~(a) and (c)]. Furthermore, the $2f$-detection of the signal indicates that these current signals are triggered by a temperature change in reservoir H. In the vertex region, the occupation numbers of both QDs can fluctuate while the occupation number becomes fixed when moving away from this region. It is thus apparent that the current changes which give rise to the clover-leaf structure originate from fluctuating occupation numbers of QD1 and QD2.
As an example, Fig.~\ref{Cartoon}~(a) shows the alignment of $\mu^{(1)}$ and $\mu^{(2)}$ with $V_\text{SD}<0$ for section 1 with $N$+1 electrons on QD1 and $M$ electrons on QD2.
Due to the ac-character of the heating current $T_\text{H}$ oscillates between the two values $T_\text{H} = T_\text{base}$ and $T_\text{H} = T_\text{max} > T_\text{base}$. The first case is shown on the left side of Fig.~\ref{Cartoon}~(a): QD1 is occupied with $N$+1 electrons, i.e. $\mu^{(1)}$ is below $\mu_\text{H}$ and therefore the electron number of QD1 is fixed at $N$+1. QD2 is occupied with $M$ electrons and the chemical potential $\mu^{(2)}$($N$+1, $M$+1) which is required to add the ($M$+1)$^{\rm th}$ electron lies outside the bias window $V_\text{SD}$. Thus, transport across QD2 is blocked. 
When the temperature in reservoir H is increased such that $T_\text{H} = T_\text{max}$ [right hand side in Fig.~\ref{Cartoon}~(a)], empty states are created below $\mu_\text{H}$ in this reservoir. This increases the charge fluctuation rate on QD1. 
However, when due to these fluctuations QD1 relaxes to the $N$-state, the energy required to add an electron to QD2 is reduced by $E_\text{C}$. The corresponding $\mu^{(2)}$($N, M$+1) is below $\mu_\text{S}$ and the current across QD2 increases [indicated by red arrows in Fig.~\ref{Cartoon}~(a)].
Since $T_\text{H}$ oscillates between $T_\text{base}$ and $T_\text{max}$, this effectively leads to a temperature driven modulation of the conductance of QD2: If $T_\text{H}$ increases, the current across QD2 increases as well. For $T_\text{H}$ at a minimum, transport is blocked. The resulting current modulation at the drain contact is then detected by the lock-in amplifier as a positive signal. 

The QD levels for section 2, with occupations ($N$+1, $M$+1), are depicted in Fig.~\ref{Cartoon}~(b). Starting again with the condition $T_\text{H} = T_\text{base}$ so that $N$+1 is fixed, we find that transport across QD2 is enabled because $\mu^{(2)}$($N$+1, $M$+1) is situated within $V_\text{SD}$. However, charge fluctuations on QD1, which increase with increasing $T_\text{H}$ [right side in Fig.~\ref{Cartoon}~(b)], tend to block transport across QD2: The corresponding $\mu^{(2)}$($N$, $M$+1) is below $\mu_\text{D}$ and thus, electrons are trapped on QD2. 
The correlation between $T_\text{H}$ and $I_\text{D}$ is now inverted compared to section 1: a temperature increase tends to block transport while small $T_\text{H}$ increase $I_\text{D}$. 
The sign of the signal in sections 3 and 4 can be understood in a similar manner: In these sections QD1 is in the $N$-state and fluctuations lead to an occasional occupation with $N$+1 electrons. This is a reversal of the situation in sections 1 and 2 because an increase of $T_\text{H}$ now causes a decrease of $I_\text{D}$ when QD2 is in the $M$-state (section 3) while it enhances $I_{\rm D}$ for QD2 exhibiting $M$+1 electrons (section 4). 
Within this picture, the explanation of the observations for a reversed voltage bias [$V_\text{SD}=100~\mu$V, cf. Fig.~\ref{Butterfly}~(b)] is also straightforward: Because a sign change of the bias voltage reverses the dc-current through QD2 this leads to an overall reversal of the observed signal.

We have performed simple model calculations to substantiate the qualitative discussion presented above.
Assuming sequential transport across QD2, the current $I_\text{D}$ can be related to the applied difference in electro-chemical potential $V_\text{SD} = \mu_S - \mu_D$ by considering Fermi-Dirac occupation statistics $f(\mu^{(2)}, T_j) = 1/(1+\text{exp}(\mu^{(2)}-\mu_j/k_{\rm B}T_j))$, $j={\rm S,D}$ in the source and the drain contact and a single resonant QD level $\mu'^{(2)} $ which is located at $\mu^{(2)} = -E_\text{C}/2 $. For $\mu_S < \mu_D$ we can then write $I_{\rm D}'(\mu^{(2)}) \propto f(-E_{\rm C}/2, T_{\rm S})\times (1-f(-E_{\rm C}/2, T_{\rm D}))$. 
The current $I_\text{D}''$ across QD2 when QD1 hosts $N$+1 electrons can be treated likewise, with $\mu^{(2)} = +E_\text{C}/2$.
The total current $I_\text{D}$ through QD2 is now the sum of $I_\text{D}'$ and $I_\text{D}''$, weighted with the appropriate probabilities of QD1 hosting $N$ or $N$+1 electrons. Thus, $I_D (\mu^{(1)},\mu^{(2)}) \propto f(\mu^{(1)}, T_{\rm H})I'_{\rm D} + (1-f(\mu^{(1)}, T_{\rm H}))I''_{\rm D}$. 

\begin{figure}[]
	\centering
		\includegraphics[width=0.7\linewidth]{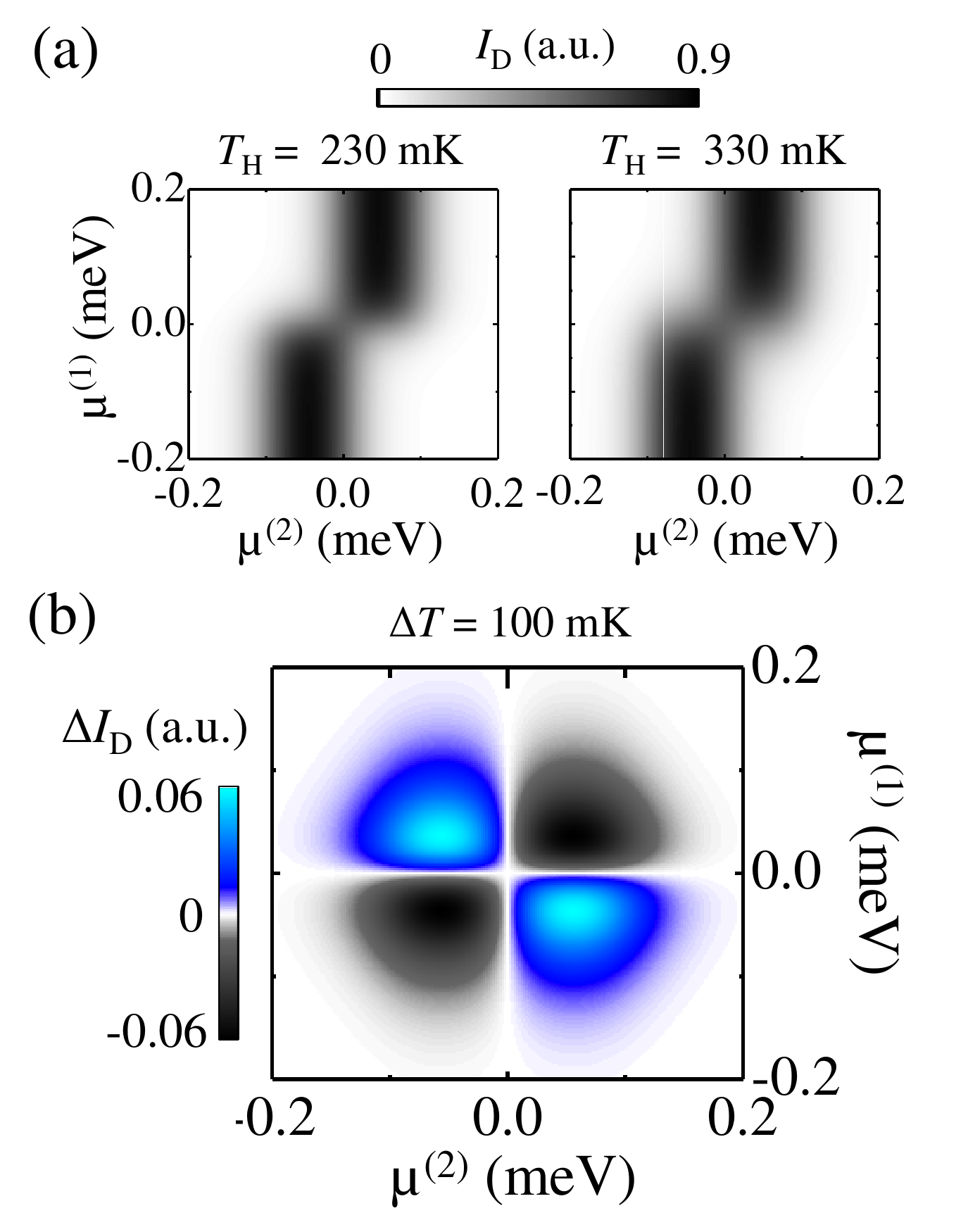}
		\caption{Model calculations for $I_\text{D}$ and $\Delta I_\text{D}$. The following parameters were used: $T_\text{S,D} = T_\text{base} = 230~\text{mK}$, $\mu_\text{S}=-\mu_\text{D} = - 50~\mu\text{eV}$, $\mu_\text{H}=0$ and $E_\text{C}=90~\mu\text{eV}$. \textbf{(a)} $I_\text{D}$ for $T_\text{H} = 230~\text{mK}$ (left) and   $T_\text{H} = 330 ~\text{mK}$ (right). \textbf{(b)} Subtraction of the figures given in (a) yields $\Delta I_\text{D} = I_\text{D}(T_\text{H} = 330~\text{mK})-I_\text{D}(T_\text{H} = 230~\text{mK})$.
		}
	\label{Model}
\end{figure}

Figure~\ref{Model} (a) compares $I_{\rm D}$ as a function of $\mu^{(1)}$ and $\mu^{(2)}$ for $T_\text{H}=T_\text{base}$ (left) and $T_\text{H}=T_\text{base}+\Delta T = T_\text{max}$ with $\Delta T = \unit[100]{mK}$ (right) while $T_\text{S,D} = T_\text{base} = \unit[230]{mK},~\Delta \mu = \unit[100]{\mu V} \text{ and } E_C = \unit[90]{\mu V}$.
As expected, the results strongly resemble the conductance stability diagram in the vertex region. However, major differences for different $T_\text{H}$ are not directly obvious. 
In order to model our experiment we subtract the calculated data sets in Fig.~\ref{Model}~(a) from each other such that we obtain $\Delta I_\text{D} = I_\text{D}(T_\text{max})-I_\text{D}(T_\text{base})$, which corresponds to the change in current through QD2 due to a change of $T_\text{H}$ by $\Delta T$. The result is given in Fig.~\ref{Model}~(b). Evidently, the clover-leaf pattern found in the experiments, is reproduced nicely.

We point out that a similar four-leafed clover pattern has been observed in connection with Coulomb coupled double QDs previously: McClure \textit{et al.} \cite{McClure2007} have reported on experiments addressing the cross-correlation of shot noise in such a system. There, the authors observed regions of positive and negative correlation at the stability region vertex arranged in a cloverleaf shaped pattern similar to the one discussed here. 
The underlying mechanism is actually closely related to the one active in our experiments:
Negatively correlated shot noise indicates that charge fluctuations of one QD tend to suppress fluctuations on the other one (and vice versa). Correspondingly, those are the configurations for which we observe a reduced current through QD2 if the temperature in reservoir H is increased. A positive correlation implies that occupation fluctuations tend to occur simultaneously on both dots. Thus, we observe an enhancement of the current through QD2 with temperature in those regions.

In order to estimate the gating range of our device we analyze the data shown in Figs.~\ref{Butterfly}~(a) and \ref{Butterfly}~(c): Using the $G$ of QD2 at those configurations for which a maximal $\Delta I_\text{D}= \unit[18]{pA}$ is observed ($G= \unit[0.09]{e^2/h}$) we calculate the drain current for $V_\text{SD} = \unit[100]{\mu V}$ and $T_\text{H} = T_\text{base}$, which gives $I_\text{D} = \unit[360]{pA}$. Relating this current to $\Delta I_{\rm D}$ then yields a gating amplitude of 5\%. Although this ratio is rather small, it can be strongly enhanced by tuning the parameters $E_\text{C}$ and $k_{\rm B} \Delta T$.
 
Finally, we note that the thermal gating effect presented here could be used, e.g., to monitor carrier heating in quantum circuits. Furthermore, it could be utilized to also manipulate heat flow across QD2: Since the thermal conductance $\kappa$ of a QD as a function of $\mu$ usually follows the Wiedemann-Franz rule and thus has a similar line shape as the conductance \cite{Guttman1995}, the mechanism presented here would allow gating of heat currents to be accomplished, thus suggesting a route to realizing a QD-based all thermal transistor.

The authors gratefully acknowledge discussions with B. Sothmann, R. S\'anchez and M. B\"uttiker.
This work has been financially supported by the DFG (SPP1386).

\end{document}